\documentclass[12pt]{article}
\input epsf.sty
\def\be{\begin{equation}}
\def\ee{\end{equation}}
\def\ba{\begin{align}}
\def\ea{\end{align}}
\def\beq{\begin{eqnarray}}
\def\eeq{\end{eqnarray}}

\def\p{\partial}

\def\ops[#1]{\p_{#1} e^{-2\phi}}

\def\eq[#1]{equation (\ref {eq:#1})}
\def\Eq[#1]{Equation (\ref {eq:#1})}
\def\e[#1]{\ref {eq:#1}}
\def\at[#1]{| _{#1}}

\let\oldpercent\%\renewcommand{\%}{\scalebox{0.85}{\oldpercent}}

\begin{document}

\baselineskip=18pt

\begin{center}
{\Large \bf{On string theory on deformed BTZ\\ and $T\bar T+J\bar T+T\bar J$}}

\vspace{10mm}

\textit{Amit Giveon and Daniel Vainshtein}
\break

Racah Institute of Physics, The Hebrew University \\
Jerusalem, 91904, Israel

\vspace{6mm}

giveon@mail.huji.ac.il\\
daniel.vainshtein@mail.huji.ac.il

\end{center}


\vspace{10mm}

\begin{abstract}

Aspects of superstring theory on black-strings backgrounds, corresponding to deformed BTZ black holes, formed
near $k$ $NS5$ branes by $p$ fundamental strings, and single-trace $T\bar T+J\bar T+T\bar J$ holography, are presented.
It is shown that for a particular asymptotic value of the $B$-field,
the excitation energy of a long string plus its contribution to the energy of the black hole
is that in $T\bar T+J\bar T+T\bar J$ deformed $CFT_2$ with $c=6k$.
The excitation energy of a winding $w>1$ long string plus its contribution to the background,
a $w/p$ fraction of the black-hole energy,
evolves according to that in a $Z_w$ twisted sector of a $p$-fold symmetric product of the latter.

\end{abstract}
\vspace{10mm}

\section{Introduction}

The perturbative string theory on $AdS_3$ ($\times\,$internal-space)
is a tractable framework in the context of the $AdS_3/CFT_2$ correspondence
(see e.g.~\cite{Aharony:1999ti}, for a review).
The worldsheet sigma model on $AdS_3$ is a solvable two-dimensional CFT;
it is a WZW model on the group manifold $SL(2,R)$, with $k$ units of $H=dB$ flux.
Consequently, many properties~\footnote{Some of which are recalled below.} of its holographic $CFT_2$ dual
can be derived by standard perturbative string theory tools.

An example, is superstring theory on $AdS_3$,
constructed near $k$ $NS5$ branes on $S^1_x\times T^4$ and $p$ fundamental strings ($F1$) on $S^1_x$,
in which case the internal space is $SU(2)_k\times T^4$.
For concreteness, we will focus on this example,
though our study is manifestly applicable to superstring theory on any $AdS_3\times S^1\times{\cal N}$.~\footnote{As well as
more general cases, e.g. those in~\cite{Giveon:1999jg,Giveon:2003ku}.}

The spectrum of perturbative strings excitations in string theory on $AdS_3$ consists of long and short strings,
which amount to states in the continuous and discrete series representations of the $SL(2,R)_k$ theory, respectively.
In addition, the spacetime $CFT_2$ has states which are dual to asymptotically $AdS_3$ geometries,
such as BTZ black holes, and consequently to strings excitations in string theory on BTZ worldsheet CFT.

It is well known,~\cite{Argurio:2000tb,Giveon:2005mi},
that the long-strings spectrum has the patten of a $p$-fold symmetric orbifold,
$({\cal M}_{6k})^p/S_p$, where the block ${\cal M}$ is a $CFT_2$ of central charge $c=6k$,
describing the target-space dynamics of a single long string.
Discrete series states are known not to fit into the symmetric product.
However, these states have measure zero in the string spectrum, so one can say
that the symmetric product describes generic states in perturbative string theory on $AdS_3$.~\footnote{See e.g.
the discussion in~\cite{Chakraborty:2023mzc}, for the present status of this picture, as well as~\cite{Balthazar:2021xeh},
for a full-fledged $CFT_2$ dual, when $k<1$, and~\cite{Eberhardt:2021vsx,Chakraborty:2025nlb},
for its proposed perturbative $CFT_2$ generalization to $k\geq 1$.}

In this note, we will consider long-strings excitations.
Winding one long strings, delta-normalizable states in $AdS_3$ with $w=1$,
are dual to states in the untwisted sector of ${\cal M}^p/S_p$.
Long-string excitations with winding $w>1$ amount to states in the $Z_w$ twisted sector of the spacetime $CFT_2$ dual,
instead.~\footnote{Prior to taking a near $F1$ limit,
$w$ amounts to winding on the same $S^1_x$
on which the $p$ strings that form the background wind (see e.g.~\cite{Chakraborty:2024ugc} with~\cite{Chakraborty:2023mzc} and below);
in the next sections, we will denote this $w$ by $w_x$
(to distinguish it from another winding number).}

A tractable extension of the $AdS_3/CFT_2$ framework is the investigation of its deformations.
Intriguingly, a class of solvable, marginal deformations of the $SL(2,R)$ worldsheet CFT,
gives rise to solvable irrelevant $CFT_2$ deformations in the boundary theory.
Accordingly, states in the spacetime $CFT_2$ that are dual to BTZ geometries, correspond in the deformed boundary theory
to deformed BTZ solutions.
The study of such deformations, starting in~\cite{Giveon:2017nie},
gave rise to an intriguing class of holographic duality,
which is referred to as ``single-trace $T\bar T$ holography'' (see e.g.~\cite{Giveon:2017myj,Apolo:2019zai,Chang:2023kkq,Chakraborty:2023mzc,Cui:2023jrb,Chakraborty:2023zdd,Giveon:2023gzh,Chakraborty:2024ugc,Chakraborty:2024mls,Giveon:2024sgz,He:2025ppz}), for reasons (some of which) we recall in the following.

The geometry of the deformed theory corresponding to the UV of the hologram
is an asymptotically flat spacetime with a linear dilaton,
whose strong-coupling regime, corresponding to the IR, is capped by a locally $AdS_3$ wall.
The long-strings spectrum in string theory on the deformed $AdS_3$ theory
has the pattern of that in a $p$-fold symmetric product whose seed, ${\cal M}$, is now a $T\bar T$ deformed $CFT_2$ with $c=6k$:
the energies of states in the $w=1$ sector evolve
via the $T\bar T$ formula of~\cite{Smirnov:2016lqw,Cavaglia:2016oda},
and those with a $w>1$ winding number evolve as in the $Z_w$ twisted sector of ${\cal M}^p/S_p$.
Consequently, the study of string theory on the deformed $AdS_3$
provides concrete properties of a hologram in asymptotically flat spacetime with a linear dilaton.

Since the long strings amount to delta-normalizable states, they have support in the entire asymptotic regime.
Hence, their energies can be obtained,~\cite{Giveon:2019fgr}, by using just the data of the flat spacetime:
its constant metric and $B$-field, and the linear-dilaton slope.
This allows one,~\cite{Chakraborty:2024ugc,Giveon:2024sgz}, to investigate the energy spectrum of long-strings excitations also
in string theory on deformed BTZ black-holes CFT, using the standard tools of perturbative string theory in flat spacetime.
The results are the following.

For a particular value of the $B$-field,~\footnote{By ``the $B$-field,'' we shall  sometimes refer to its asymptotic constant value;
this should be clear by context.} the excitation energy of a winding $w$ probe long string, $E$,
plus its contribution to the black-hole background, $w$ times the $p$-fraction of the black-hole energy, $E_{BH}$,
is in harmony with single-trace $T\bar T$ holography.
Concretely, the energy spectrum of wound strings in the deformed theory,
\be\label{etint}
E_{\rm total}\equiv E+w{E_{BH}\over p}~,
\ee
with undeformed spectrum
\be\label{udint}
{\cal E}_{\rm total}\equiv{\cal E}+w{\sqrt{\alpha' k}M_{BTZ}\over p}~,
\ee
where $M_{BTZ}$ is the BTZ black-holes mass and ${\cal E}$ the dimensionless energy of the probe string in the undeformed BTZ,
evolves as in $T\bar T$ deformed $CFT_2$ with $c=6k$, in the $w=1$ sector,
and as in the $Z_w$ twisted sector of its symmetric product, in the $w>1$ sectors~\cite{Chakraborty:2024ugc,Giveon:2024sgz}.

An extension of the above is the study of string theory on deformed $AdS_3\times S^1_y$,
initiated in~\cite{Chakraborty:2018vja,Apolo:2018qpq,Araujo:2018rho,Giveon:2019fgr,LeFloch:2019rut,Chakraborty:2019mdf}.
The deformed $AdS_3\times S^1$ geometry amounts to the sigma-model metric, $B$-field and dilaton,
obtained in a family of truly-marginal deformations of an $SL(2,R)\times U(1)$ worldsheet theory.
The resulting worldsheet theories are still solvable,
and thus allow the investigation of many properties of the two-dimensional spacetime dual.
We refer to the duality as ``single-trace $T\bar T+J\bar T+T\bar J$ holography,''
for reasons (some of which) we recall next.

The study of the spectrum of string theory on deformed $AdS_3\times S^1$ was initiated
in~\cite{Chakraborty:2018vja,Apolo:2018qpq,Chakraborty:2019mdf}.
Following~\cite{Chakraborty:2019mdf}, we shall denote the deformation parameters of the geometry by $(\lambda,\epsilon_\pm)$ below.
On the field theory side,
the study of the spectrum of $J\bar T$ deformed $CFT_2$ was initiated in~\cite{Guica:2017lia},
and that of the whole family of $T\bar T+J\bar T+T\bar J$ deformed $CFT_2$ in~\cite{LeFloch:2019rut}.

The result of~\cite{Chakraborty:2019mdf} is that the long-strings spectrum in string theory on the
$(\lambda,\epsilon_\pm)$-deformed $AdS_3\times S^1$ theory
has the same pattern as that in a $p$-fold symmetric product whose seed, ${\cal M}$, is now
a linear combination of $T\bar T$, $T\bar J$ and $J\bar T$ deformed $CFT_2$ with $c=6k$,
that has an affine $U(1)_L$ and $U(1)_R$ symmetries,
with the corresponding left and right-handed currents, $J$ and $\bar J$, respectively.
The dimensionless $T\bar T$ and $J\bar T,T\bar J$ couplings in ${\cal M}$ are the $\lambda$ and $\epsilon_\pm$
of the deformed $AdS_3\times S^1$ sigma-model geometry, respectively.
We shall denote the deformation of the $CFT_2$ by `$T\bar T+J\bar T+T\bar J$,'
regardless of the values of its irrelevant deformation parameters.

More concretely, one finds,~\cite{Chakraborty:2019mdf},
that the energies of states in the $w=1$ sector of string theory on deformed $AdS_3\times S^1$
evolve via the $T\bar T+J\bar T+T\bar J$ formula,
and those with a $w>1$ winding number evolve as in the $Z_w$ twisted sector of ${\cal M}^p/S_p$,
for any~$(\lambda,\epsilon_\pm)$.
And, more precisely,
in~\cite{Chakraborty:2018vja,Apolo:2018qpq,Chakraborty:2019mdf},
the boundary spatial direction of $AdS_3$ is compact, a.k.a. its global structure is that of a massless BTZ geometry.
Strictly speaking, the results above are thus in string theory on deformed massless BTZ$\times S^1$.

Now, for generic values of the deformation parameters, $(\lambda,\epsilon_\pm)$,
the geometry of the deformed theories interpolates between an asymptotically flat spacetime with a linear dilaton in the UV
and $AdS_3$ in the IR, as in the $T\bar T$ case above.
So, as in the $\epsilon_\pm=0$ case,
the energy spectrum of long strings in the asymptotically linear-dilaton spacetime of string theory on deformed $AdS_3\times S^1$
can be obtained,~\cite{Giveon:2019fgr}, by using just the asymptotic constant values of the metric and $B$-field, and the linear-dilaton slope.

Consequently, this allows one to investigate the energy spectrum of long-strings excitations also in string theory on deformed
{\it massive} BTZ black-holes CFT, times a circle, using the standard tools of perturbative string theory in flat spacetime.
This is the purpose of this note.

In section 2, we consider superstring theory on the deformed massless BTZ$\times S^1$ black-holes backgrounds of~\cite{Chakraborty:2019mdf},
formed in the near $k$ $NS5$ branes on $S^1_x\times S^1_y\times T^3$
with $p$ fundamental strings ($F1$) wrapping the $S^1_x$,
and re-derive its long-strings spectrum from the asymptotic flat spacetime (with linear dilaton) data.

In section 3, we turn to the massive BTZ black-hole case, $M_{BTZ}>0$.
The question we address is:
what is the asymptotic value of the $B$-field required for the long-strings spectrum
in $(\lambda,\epsilon_\pm)$-deformed massive BTZ$\times S^1_y$
to be in harmony with single-trace $T\bar T+J\bar T+T\bar J$ holography?
We find the answer:
for a particular value of the $B$-field (in eq.~(\ref{answer})),
similar to~(\ref{etint}),(\ref{udint}),
the excitation energy of a winding $w$ probe long string
plus its contribution to the black-hole background, a $w/p$-fraction of the black-hole energy
(the $E_{\rm total}$ in~(\ref{especb})--(\ref{calet})),
evolves in harmony with single-trace $T\bar T+J\bar T+T\bar J$ holography,
for any $(\lambda,\epsilon_\pm)$ and $M_{BTZ}$.

Finally, in section 4, we summaries the main results,
and in the appendices, we present some technical details.

\section{Long strings in $(\lambda,\epsilon_\pm)$-deformed massless BTZ}

The spectrum of long strings in string theory on $(\lambda,\epsilon_\pm)$-deformed {\it massless} BTZ worldsheet CFT
was inspected using different methods in~\cite{Chakraborty:2019mdf,Apolo:2021wcn,Hashimoto:2019wct}.
In this section, as a warmup, setting the stage, and providing our tools,
we re-consider the investigation in superstring theory on such
deformed $M_{BTZ}=0$ black holes times a circle.~\footnote{From another perspective.}

For concreteness, we present the discussion for superstring theory on black-strings backgrounds,
corresponding to deformed BTZ black holes formed near $k$ NS5 branes on $S^1_x\times S^1_y\times T^3$,
by $p$ fundamental strings ($F1$), wrapping the $x$-circle.

Following~\cite{Chakraborty:2019mdf},~\footnote{With slight modifications of notation and w.l.g. choices.}
the sigma-model action, $S(\lambda,\epsilon_\pm)$, and dilaton, $\Phi$,
of the $(\lambda,\epsilon_\pm)$-deformed massless BTZ worldsheet CFT are~\footnote{We ignore the internal space,
${\cal N}=SU(2)_k\times T^3$, and note that our study extends to any ${\cal N}$.}
\be\label{action}
S(\lambda,\epsilon_\pm)={1\over 2\pi\alpha'}\int d^2z\left[\partial\phi\bar\partial\phi+h\left(\partial\gamma_+\bar\partial\gamma_-
+2\epsilon_+\partial y\bar\partial\gamma_-+2\epsilon_-\partial\gamma_+\bar\partial y+f^{-1}\partial y\bar\partial y\right)\right],
\ee
where~\footnote{To simplify the discussion below, we restrict to the $(\lambda,\epsilon_\pm)$-subspace in which
$$\lambda-4\epsilon_-\epsilon_+>0~,$$
such that the $r_x$ below is real (and positive). We'll comment later on the continuation to $\lambda-4\epsilon_-\epsilon_+<0$.}
\be\label{gammas}
\gamma_\pm\equiv{1\over r_x}(x\pm t)~,\qquad r_x={1\over\sqrt{\lambda-4\epsilon_-\epsilon_+}}~,
\ee
\be\label{hf}
h^{-1}={1\over r_x^2}+e^{-2\phi/r_5}~,\qquad f^{-1}=\lambda+e^{-2\phi/r_5}~,\qquad r_5\equiv\sqrt{\alpha' k}~,
\ee
\be\label{rxry}
x\sim x+2\pi R_x~,\quad R_x=\sqrt{\alpha'}r_x~,\qquad y\sim y+2\pi R_y~,\quad R_y=\sqrt{\alpha'}r_y~,
\ee
and~\footnote{Note that without loss of generality, we chose the periodicity $${\gamma_\pm\over\sqrt{\alpha'}}\sim{\gamma_\pm\over\sqrt{\alpha'}}+2\pi~.$$}
\be\label{dil}
e^{2\Phi}=g_s^2e^{-2\phi/r_5}h~,\qquad g_s^2\equiv e^{2\Phi(\phi\to -\infty)}={kv\over p}~,\quad
v\equiv{{\rm Volume}(S^1_y\times T^3)\over(2\pi\sqrt{\alpha'})^4}~.
\ee

As one approaches the asymptotically flat spacetime, $R_\phi\times R_t\times S^1_x\times S^1_y$ with linear dilaton, at $\phi\to\infty$,
the metric, $B$-field and dilaton approach the values
\be\label{dsto}
ds^2\to -dt^2+dx^2+2r_x(\epsilon_-+\epsilon_+)dxdy
+2r_x(\epsilon_--\epsilon_+)dtdy
+r_x^2 \lambda dy^2+d\phi^2~,
\ee
\be\label{bto}
B_{tx}\to 1~,\qquad B_{ty}\to r_x(\epsilon_-+\epsilon_+)~,\qquad B_{xy}\to r_x(\epsilon_--\epsilon_+)~,
\ee
and
\be\label{dilto}
\Phi\to -{\phi\over r_5}+const~.
\ee

A large class of observables in the (NS,NS) sector of superstring theory on the deformed BTZ black holes
in~(\ref{action})--(\ref{dil}) is given by vertex operators
in the $(-1,-1)$ picture, whose behavior in the asymptotically linear-dilaton regime,~(\ref{dsto})--(\ref{dilto}), is
\be\label{vb}
V_{phys}\to e^{-\varphi-\bar\varphi}V_{N_L,N_R}e^{-iE t}e^{ip_Lx_L+ip_Rx_R}e^{iq_Ly_L+iq_Ry_R}e^{2j\phi/r_5}~,
\ee
where~\footnote{We follow the conventions of eqs. (12),(13) in~\cite{Chakraborty:2024ugc}; the $w,n$ there (and in the introduction),
are denoted here by $w_x,n_x$ (to distinguish them from the extra quantum numbers, $w_y,n_y$).}
\be\label{ppqq}
(p_L,p_R)=\left({w_xR_x\over\alpha'}+{n_x\over R_x},{w_xR_x\over\alpha'}-{n_x\over R_x}\right)~,\qquad
(q_L,q_R)=\left({w_yR_y\over\alpha'}+{n_y\over R_y},{w_yR_y\over\alpha'}-{n_y\over R_y}\right)~.
\ee
Each of these operators amounts to a string with windings $w_{x,y}$ and momentum $n_{x,y}$ on the asymptotic circles, $S^1_{x,y}$,
moving with momentum governed by the quantum number $j$ in the radial direction,
in a particular state of transverse left and right-handed levels, $N_{L,R}$; its energy is $E$.

The on-shell condition implies that
\be\label{onshellb}
\Delta+\bar\Delta-{2j(j+1)\over k}+N_L+N_R-1=0~,
\ee
where $\Delta+\bar\Delta$ is the scaling dimension of the operator $e^{-iE t}e^{ip_Lx_L+ip_Rx_R}e^{iq_Ly_L+iq_Ry_R}$.
For canonically normalized time, $t$, times two circles with dimensionless radii $r_{x,y}$,~(\ref{rxry}),
manipulations in~\cite{Giveon:1994fu}, adapted to time times two circles, give that~\footnote{See eq. (2.4.11) in~\cite{Giveon:1994fu}.}
\be\label{dpd}
\Delta+\bar\Delta={1\over 2}\left(n_i(g^{-1})^{ij}n_j+m^i(g-bg^{-1}b)_{ij}m^j+2m^ib_{ik}(g^{-1})^{kj}n_j\right)~,
\ee
where in our case,~(\ref{dsto})--(\ref{dilto}),
\be\label{gij}
g_{ij}=\pmatrix{
 -1&0&r_xr_y(\epsilon_--\epsilon_+)\cr
 0&r_x^2&r_x^2r_y(\epsilon_-+\epsilon_+)\cr
 r_xr_y(\epsilon_--\epsilon_+)&r_x^2r_y(\epsilon_-+\epsilon_+)&r_x^2r_y^2\lambda
},
\ee
\be\label{bij}
b_{ij}=\pmatrix{
0 & -r_x & -r_xr_y(\epsilon_-+\epsilon_+)\cr
r_x& 0& -r_x^2r_y(\epsilon_--\epsilon_+)\cr
r_xr_y(\epsilon_-+\epsilon_+)&r_x^2r_y(\epsilon_--\epsilon_+)&0
},
\ee
and~\footnote{Note that we chose the overall sign of $b_{ij}$ (relative to $B_{ij}$), and of $m^i$ (relative to $w_{x,y}$), s.t.
$$ b_{ij}=\pmatrix{
0 & -r_xB_{tx} & -r_yB_{ty}\cr
r_xB_{tx}& 0& -r_xr_yB_{xy}\cr
r_yB_{ty}&r_xr_yB_{xy}&0
},$$
and that the spectrum,~(\ref{dpd}), is invariant under $b_{ij}\to -b_{ij}$ (with $m^i\to -m^i$).}
\be\label{niwi}
n_i=(e,n_x,n_y)~,\qquad m^i=(0,w_x,w_y)~,\qquad i,j=t,x,y~,
\ee
for dimensionless energy,
\be\label{eee}
e\equiv\sqrt{\alpha'}E~,
\ee
from which one finds that~\footnote{In appendix A,
we present more steps of the algebra, for the $\epsilon_-=\epsilon_+$ case.}
\beq\label{finds}
\Delta+\bar\Delta&=&
-\frac{1}{2}\Psi r_x^2e^2
\\
&+&\left(-w_{x}+\left(\epsilon_{+}^{2}-\epsilon_{-}^{2}\right)n_{x}
+\left(\epsilon_{-}-\epsilon_{+}\right)\frac{n_{y}}{r_{y}}-\left(\epsilon_{-}+\epsilon_{+}\right)w_{y}r_{y}\right)r_{x}e
\nonumber\\
&+&\frac{1}{2}\left(\lambda+\left(\epsilon_{-}-\epsilon_{+}\right){}^{2}\right)n_{x}^{2}
+\left(\left(\epsilon_{-}-\epsilon_{+}\right)w_{y}r_{y}-\left(\epsilon_{-}+\epsilon_{+}\right)\frac{n_{y}}{r_{y}}\right)n_{x}
\nonumber\\
&+&\frac{1}{2}\left(\frac{n_{y}^{2}}{r_{y}^{2}}+r_{y}^{2}w_{y}^{2}\right),\nonumber
\eeq
where
\be\label{defpsi}
\Psi\equiv\lambda-(\epsilon_-+\epsilon_+)^2~,
\ee
and
\be\label{dmd}
\qquad\Delta-\bar\Delta=n_xw_x+n_yw_y~.
\ee

The $(\Delta,\bar\Delta)$ in (\ref{dpd})--(\ref{dmd}) are identical to those in (5.27) of~\cite{Chakraborty:2019mdf} (CGK)
with the map~\footnote{Which
follows from the factor $R_{CGK}$ difference in our canonical choice (w.l.g.) of the $\gamma_1$ periodicity
(versus the generic, redundant one in~\cite{Chakraborty:2019mdf}).}
\be\label{mapcgkb}
(ER)_{CGK}=r_x e~,\qquad \hat{\lambda}_{CGK}=2\lambda~,\qquad(\hat\epsilon_\pm)_{CGK}=\sqrt{2}\epsilon_\mp~,
\ee
and
\be\label{qqcgkb}
(q_L,q_R)_{CGK}={1\over\sqrt{2}}\left({n_y\over r_y}-w_yr_y,{n_y\over r_y}+w_yr_y\right)~,
\ee
instead of~(\ref{ppqq}),~\footnote{In terms of $(q_L,q_R)$, the map is thus $(q_L,q_R)_{CGK}=\sqrt{\alpha'\over 2}(-q_R,q_L)$,
due to $\sqrt{2/\alpha'}$ and minus factors in~(\ref{ppqq}) versus $(q_L,q_R)_{CGK}$ in~(\ref{qqcgkb});
the interchange of the left and right-handed $\epsilon_\pm$ in~(\ref{mapcgkb})
is in accordance with the interchange of $q_{L,R}$.}
as well as $(w,n)_{CGK}\equiv(w_x,n_x)$.

Now, in terms of $E$, $R_x$ and $(q_L,q_R)$ in~(\ref{eee}), (\ref{rxry}) and~(\ref{ppqq}), respectively,
and with the $(\Delta,\bar\Delta)$ in equations~(\ref{finds})--(\ref{dmd}),
the on-shell condition~(\ref{onshellb}) takes the form
\be\label{redpd}
a\left(ER_x\right)^{2}+b(ER_x)+c=0~,
\ee
with
\be\label{a}
a=-\frac{\Psi}{2}~,
\ee
\be\label{bb}
b=-w_{x}+\left(\epsilon_{+}^{2}-\epsilon_{-}^{2}\right)n_{x}
-\sqrt{\alpha'}\left(q_{R}\epsilon_{-}+q_{L}\epsilon_{+}\right)~,
\ee
and
\be\label{c}
c=\frac{1}{2}\left(\lambda+\left(\epsilon_{-}-\epsilon_{+}\right){}^{2}\right)n_{x}^{2}
+\sqrt{\alpha'}\left(q_{R}\epsilon_{-}-q_{L}\epsilon_{+}\right)n_{x}+w_x{\cal E}~,
\ee
where~\footnote{We restrict to strings with positive $w_x$ and $j=-\frac{1}{2}+is$, $s\in R$, below;
see e.g. the discussion in~\cite{Giveon:2017myj,Chakraborty:2019mdf,Chakraborty:2024mls} regarding negative versus positive $w_x$ states,
and the discussion in e.g.~\cite{Chakraborty:2023mzc,Chakraborty:2023zdd,Chakraborty:2024mls} concerning discrete states versus those
in the continuous representations, which we consider here.}
\be\label{lsbtz}
{\cal E}={1\over w_x}\left[-{2j(j+1)\over k}+\frac{\alpha'}{4}\left(q_{L}^{2}+q_{R}^{2}\right)+N_L+N_R-1\right]~,
\ee
with ${\cal E}$ being the dimensionless energy carried by the string excitation~(\ref{vb}) in the undeformed BTZ theory,
as explained in~\cite{Chakraborty:2018vja,Chakraborty:2019mdf}.~\footnote{The gist:
for long strings, for each winding number, $w_x=1,2,3,\dots$, with all other quantum numbers and radial momentum fixed,
we consider the flow of energies, $E(\lambda,\epsilon_\pm)$, and recall (5.37) in~\cite{Chakraborty:2018vja},
that long strings in massless BTZ satisfy eq.~(\ref{lsbtz}); in~\cite{Chakraborty:2018vja,Chakraborty:2019mdf},
${\cal E}$ was denoted by $h_w+\bar h_w-{kw\over 2}$, using the $AdS_3/CFT_2$ dictionary,
with the $w$ there being the $w_x$ here.}
Indeed, in the $\lambda,\epsilon_\pm\to 0$ limit, $a\to 0$, $b\to w_x$ and $c\to w_x{\cal E}$,
and thus~\footnote{In~\cite{Chakraborty:2024ugc,Giveon:2024sgz}, $E$ was denoted by $E_w$ and/or $E_{w,\rm string}$
and ${\cal E}$ by ${\cal E}_w$ and/or ${\cal E}_{w,\rm string}$, correspondingly; the $w$ and $R$ there are the $w_x$ and $R_x$ here,
respectively.}
\be\label{limitb}
\lim_{\lambda,\epsilon_\pm\to 0}ER_x={\cal E}~.
\ee
Finally, recall,~\cite{Giveon:2017nie,Giveon:2017myj,Chakraborty:2019mdf,Chakraborty:2024ugc},
that the energy spectrum,~\footnote{The branch of the solutions of~(\ref{redpd}) in~(\ref{espec}) is the one which satisfies~(\ref{limitb}).}
\be\label{espec}
ER_x={1\over 2a}\left(-b-\sqrt{b^2-4ac}\right)~,
\ee
with the $a,b,c$ in (\ref{a}),(\ref{bb}),(\ref{c}), is in harmony with single-trace $T\bar T+J\bar T+T\bar J$ holography,
where the geometrical parameters, $\lambda$ and $\epsilon_\pm$ in the worldsheet sigma-model~(\ref{action})--(\ref{hf}),
are (proportional to) the dimensionless $T\bar T$ and $J\bar T,T\bar J$ couplings, respectively,
in the boundary spacetime theory.~\footnote{For
the recent status of single-trace $T\bar T$ holography, see e.g. the discussion in~\cite{Chakraborty:2023mzc},
and its followups,~\cite{Chakraborty:2024ugc,Chakraborty:2023zdd};
by ``single-trace $T\bar T+J\bar T+T\bar J$ holography,''
we refer to the manifest generalization of the single-trace $T\bar T$ holography conjecture
(along the lines of~\cite{Chakraborty:2019mdf},
whose gist is in the introduction).}

\section{Long strings in $(\lambda,\epsilon_\pm)$-deformed massive BTZ}

We are now ready to investigate the energy spectrum of long strings in superstring theory on $(\lambda,\epsilon_\pm)$-deformed BTZ black holes
with mass $M_{BTZ}$,~\footnote{For simplicity, we present the non-rotating, chargeless case, $J_{BTZ}=Q_{L,R}=0$; we verified that the generalization of the results in this note to any rotating, charged black strings is straightforward.}
in the framework of single-trace $T\bar T+J\bar T+T\bar J$ holography.
Such black-string solutions amount to excitations of the sigma-model background in (\ref{action})--(\ref{dil}),
which do not affect the asymptotic behavior of its metric,~(\ref{dsto}), and dilaton~(\ref{dilto}).~\footnote{Such
solutions appear in~\cite{Apolo:2021wcn} (AS) for any $M_{BTZ}$, $J_{BTZ}$ and $Q_{L,R}$; the map of the parameters in the AS solutions to ours is subtle (see~\cite{Giveon:2024sgz} for the map in the $\epsilon_\pm=0$ case),
but it's not needed here, since we'll not use any of the properties of the solutions, beyond their asymptotic behavior.}

Since $H=dB$, adding a constant to the $B$-field of a solution is still a solution,~\footnote{It is not clear to us though
which choices give rise to consistent worldsheet CFT.}
with the same asymptotic metric and dilaton.
One is thus left with the freedom to modify the asymptotic value of the $B$-field of the black-string solutions,
as a function of $(\lambda,\epsilon_\pm)$ and $M_{BTZ}$,
with the requirement that $B_{tx},B_{ty},B_{xy}\to$ their values in equation~(\ref{bto}), when $M_{BTZ}\to 0$.

The question addressed in this section is:
{\it what is the asymptotic value required to be approached by the $B$-field, such that
the long-strings spectrum in superstring theory on such black-strings worldsheet CFTs
is in harmony with single-trace $T\bar T+J\bar T+T\bar J$ holography?}~\footnote{Needless to say that, once such a value is found,
one can adjust the constant piece of the $B$-field, accordingly.}

The answer is:
\beq\label{answer}
&&B_{tx}\to 1+{1\over\Psi r_x^2}\left(-1+\sqrt{1+2\Psi r_5 M_{BTZ}/p}\right),\\
&&B_{xy}\to r_x(\epsilon_--\epsilon_+)~,\quad B_{ty}\to r_x(\epsilon_-+\epsilon_+)~,\nonumber
\eeq
where $\Psi$ and $r_x$ are given in terms of $(\lambda,\epsilon_\pm)$ in~(\ref{defpsi}) and~(\ref{gammas}),
respectively.

A few comments are in order:
\begin{itemize}
\item
Note that
\be\label{mbtx}
(-1+B_{tx})r_x={\sqrt{\alpha'}E_{BH}\over p}~,
\ee
where~\footnote{The relation~(\ref{mbtx}) is satisfied also for general $J_{BTZ}$ and $Q_{L,R}$.}
\be\label{ebh}
E_{BH}={p\over\Psi R_x}\left(-1+\sqrt{1+2\Psi r_5 M_{BTZ}/p}\right)
\ee
is the deformed BTZ black-hole energy above extremality.~\footnote{The asymptotic value of the $B$-field in~(\ref{answer}),
which in particular depends on the black-holes mass in a certain way,~(\ref{mbtx}),
differs from that in~\cite{Apolo:2021wcn} (which is independent of the BTZ mass),
apart for the massless case; the difference between the two is thus~(\ref{mbtx}) with (\ref{ebh}).}
\item
Equation (\ref{ebh}) can be derived either in the covariant phase-space formalism,~\cite{Apolo:2021wcn},
and/or in the ADM formalism.~\footnote{It can be done e.g. by extending the analysis in appendix C of \cite{Giveon:2005mi} also to the more general black strings of~\cite{Apolo:2021wcn}; we'll not present the details here.}
Namely, it can be shown~\footnote{Since the asymptotic value of $G_{ty}$ in~(\ref{dsto}) behaves like $\epsilon_+-\epsilon_-$, the standard ADM formalism applies strictly when $\epsilon_+=\epsilon_-$; nevertheless, we assume~(\ref{ebh}) also for $\epsilon_+\neq\epsilon_-$ (see section 6.3 of~\cite{Apolo:2021wcn}).} that the black strings in the $(\lambda,\epsilon_\pm)$-deformed BTZ space (with $J_{BTZ}=Q_{L,R}=0$) have an ADM mass
\be\label{mbh}
M_{BH}=E_{BH}+{p\over\Psi R_x}~.
\ee
The constant shift (for fixed values of $\lambda,\epsilon_\pm$ and $p$) between the ADM mass, $M_{BH}(r_5M_{BTZ})$, and the energy $E_{BH}(r_5M_{BTZ})$ in eqs.~(\ref{ebh}),~(\ref{mbh}),
\be\label{intc}
C(\lambda,\epsilon_\pm;p)\equiv{p\over\Psi R_x}={pR_x\over\alpha'}{1\over 1-r_x^2(\epsilon_+-\epsilon_-)^2}
={pR_x\over\alpha'}\left(1-{(\epsilon_+-\epsilon_-)^2\over\Psi+(\epsilon_+-\epsilon_-)^2}\right)^{-1},
\ee
is fixed by requiring that the energy above the ground state of the string-fivebrane system, whose $\epsilon_\pm$ deformations form the black-strings backgrounds, vanish in the extremal case, namely, that $E_{BH}=0$ when $M_{BTZ}=0$, for any $\lambda$ and $\epsilon_\pm$.~\footnote{See e.g.~\cite{Chakraborty:2023zdd} for more details (in the $\epsilon_\pm=0$ case), which we'll not repeat; they apply also to the $\epsilon_\pm$-deformed systems here.}~\footnote{In the derivation of the black-strings mass by the covariant phase-space formalism, as was done in~\cite{Apolo:2019zai,Apolo:2021wcn,Chang:2023kkq}, $C$ is an integration constant ambiguity, which may be a function of $\lambda,\epsilon_\pm$ and $p$ (but not of $r_5M_{BTZ}$), appearing in the conserved charge (see e.g. (4.26) in~\cite{Chang:2023kkq}).}
\item
Note that while in the left-right symmetric case, $\epsilon_-=\epsilon_+$, the ground-state energy of the $\epsilon$-deformed string-fivebrane system, $C=pR_x/\alpha'$ in~(\ref{intc}), is identical to that in the undeformed system, this is modified when left-right symmetry is broken in the $\epsilon_\pm$-deformed system, a.k.a. if $\epsilon_+-\epsilon_-\neq 0$ in~(\ref{intc}).
\item
The black-strings entropy, $S_{BH}$, is fixed in the deformation space. Concretely,
\be\label{entropy}
S_{BH}=2\pi\sqrt{2kpr_5M_{BTZ}}=4\pi p\sqrt{c\over 12}\sqrt{{\cal E}_{BH}/p}=S_{BTZ}~,
\ee
with
\be\label{csixk}
c=6k~,
\ee
and
\be\label{chbh}
{\cal E}_{BH}=\lim_{\lambda,\epsilon_\pm\to 0}E_{BH}R_x=r_5M_{BTZ}~,
\ee
for all $\lambda,\epsilon_\pm$.~\footnote{The calculation of the entropy is standard; again, it can be done e.g. by extending the analysis in appendix C of~\cite{Giveon:2005mi} also to the more general black strings of~\cite{Apolo:2021wcn}, and we'll not present the details here.}
\item
The black-strings entropy $S_{BH}$ in (\ref{entropy}) and energy $E_{BH}$ in (\ref{ebh}) are in harmony with single-trace $T\bar T+J\bar T+T\bar J$ holography.
Namely (recall the introduction),
their pattern is the same as that of states in a $p$-fold symmetric product of a $T\bar T+J\bar T+T\bar J$ deformed $CFT_2$ with central charge (\ref{csixk}), whose energy is equally split among the $p$ different copies of the seed,
and with their initial energy being related to the $p$th fraction of the BTZ mass \`a la $AdS_3/CFT_2$ dictionary, (\ref{chbh}).
Concretely, the $p$th fraction of the entropy, $S_{BH}/p$, is fixed in the $(\lambda,\epsilon_\pm)$-deformed space, and is equal to that of spinless states with dimensionless energy ${\cal E}_{BH}$ in a $CFT_2$ with $c=6k$ (the one in (\ref{entropy}), (\ref{csixk})) on a cylinder with radius $R_x$, while the $p$th fraction of the energy, $E_{BH}R_x/p$, evolves via the RHS of eq.~(\ref{espec}) with the $a,b,c$ in~(\ref{a}),(\ref{bb}),(\ref{c}), respectively,
having $w_x=1$, $n_x=q_{L,R}=0$, and with the ${\cal E}$ in~(\ref{c}) being replaced by the $p$th fraction of ${\cal E}_{BH}$, a.k.a. the $p$th fraction of the BTZ mass (in units of the curvature length, $r_5=\sqrt{\alpha' k}$), ${\cal E}_{BH}/p=r_5M_{BTZ}/p$, (\ref{chbh}).
\end{itemize}

Proof of~(\ref{answer}):

First, note that~(\ref{mbtx}) means that
\be\label{bijm}
b_{ij}=\pmatrix{
0 & -r_x-e_{BH}/p & -r_xr_y(\epsilon_-+\epsilon_+)\cr
r_x+e_{BH}/p& 0& -r_x^2r_y(\epsilon_--\epsilon_+)\cr
r_xr_y(\epsilon_-+\epsilon_+)&r_x^2r_y(\epsilon_--\epsilon_+)&0
},
\ee
where
\be\label{ebhm}
e_{BH}\equiv\sqrt{\alpha'}E_{BH}~,
\ee
instead of~(\ref{bij}).~\footnote{Which is a special case of~(\ref{ebhm}), of course, with $e_{BH}=0$.}

Next, applying the machinery of the previous section,
for a long string excitation that amounts to~(\ref{vb}),
in flat spacetime with the metric and $B$-field in~(\ref{gij}) and~(\ref{bijm}), respectively, and the linear dilaton in~(\ref{dilto}),
one finds a quadratic equation, whose solution can be written in the form:
\be\label{especb}
E_{\rm total}R_x={1\over 2a}\left(-b-\sqrt{b^2-4ac}\right)~,
\ee
with
\be\label{etotal}
E_{\rm total}R_x=ER_x+w_x{E_{BH}R_x\over p}~,
\ee
where $E_{BH}$ is given in~(\ref{ebh}),
with the $a,b$ in~(\ref{a}),(\ref{bb}), respectively, and
\be\label{cb}
c=\frac{1}{2}\left(\lambda+\left(\epsilon_{-}-\epsilon_{+}\right){}^{2}\right)n_{x}^{2}
+\sqrt{\alpha'}\left(q_{R}\epsilon_{-}-q_{L}\epsilon_{+}\right)n_{x}+w_x{\cal E}_{\rm total}~,
\ee
where
\be\label{lsbtzb}
{\cal E}_{\rm total}={1\over w_x}\left[-{2j(j+1)\over k}+\frac{\alpha'}{4}\left(q_{L}^{2}+q_{R}^{2}\right)+N_L+N_R-1\right]~,
\ee
is satisfying
\be\label{calet}
{\cal E}_{\rm total}=\lim_{\lambda,\epsilon_\pm\to 0}E_{\rm total}R_x={\cal E}+w_x{r_5 M_{BTZ}\over p}~,
\ee
with ${\cal E}$ being the dimensionless energy carried by the string excitation~(\ref{vb}) in the undeformed BTZ theory
(see the previous section).~\footnote{The equality
$$
{\cal E}={1\over w_x}\left[-{2j(j+1)\over k}+\frac{\alpha'}{4}\left(q_{L}^{2}+q_{R}^{2}\right)+N_L+N_R-1\right]
-w_x{r_5 M_{BTZ}\over p}~,
$$
which follows from the RHS of~(\ref{calet}) being equal to the RHS of~(\ref{lsbtzb}),
is in agreement with the discussion around (2.32)--(2.34) in~\cite{Martinec:2023plo};
see~\cite{Chakraborty:2024ugc,Giveon:2024sgz}, for more details on the harmony with~\cite{Martinec:2023plo}.}

The result in~(\ref{especb})--(\ref{calet}) is precisely the one required in single-trace $T\bar T+J\bar T+T\bar J$ holography
(see the introduction and summary): in the $w_x=1$ sector, the total energy evolves as in a
$T\bar T+J\bar T+T\bar J$ deformed $CFT_2$,~\footnote{With central charge $6k$ (see the introduction).}
and in the $w\equiv w_x>1$ sector, $E_{\rm total}(\lambda,\epsilon_\pm)$
evolves as the energies in a $Z_w$ twisted sector of a $p$-fold symmetric product of the former.

Hence, this completes the proof.~\footnote{For illustration of the calculations leading to the results above,
in appendix B, we present more steps of the algebra, in the $n_x=q_{L,R}=0$ case (for simplicity).}

A few comments are in order:

\begin{enumerate}
\item
Although the calculations were presented in the regime of the $(\lambda,\epsilon_\pm)$-space where $\lambda-4\epsilon_-\epsilon_+>0$,
the results for the spectrum thus obtained
(in the way presented in~(\ref{especb}) via the $a,b$ and $c$ in~(\ref{a}),(\ref{bb}) and~(\ref{cb}), respectively),
apply to any $(\lambda,\epsilon_\pm)$.~\footnote{A careful consideration of the $\lambda-4\epsilon_-\epsilon_+<0$ cases
can be done e.g. similar to what is presented in the $\epsilon_\pm=0$ and $\lambda<0$ case in~\cite{Giveon:2024sgz};
we shall not present the details here.}
\item
One can show~\footnote{E.g. by rewriting the $E_{\rm total}R_x$ of~(\ref{especb}) in the form
$$E_{\rm total}R_x=n_x+{1\over 2A}\left(-B-\sqrt{B^2-4AC}\right),$$
as in (5.31) of~\cite{Chakraborty:2019mdf},
and noticing that $A=a=-\Psi/2$,~(\ref{a}), and that the $C$ thus obtained
is non-negative in the Ramond sector of a unitary $CFT_2$ (note that the $C$ thus obtained differs from that
in~(5.32) of~\cite{Chakraborty:2019mdf} just by adding to it the contribution of the excited long string in the undeformed theory
to the BTZ black-hole energy, which is positive).}
that the long-strings excitations in such black-string backgrounds
are free of complex energies iff
\be\label{psipos}
\Psi\geq 0~,
\ee
where $\Psi$ is defined in~(\ref{defpsi}).
\item
On the other hand, in the $(\lambda,\epsilon_\pm)$-subspace of `negatively deformed' theories, a.k.a. those with
\be\label{psineg}
\Psi<0~,
\ee
we note the following property.
Equation~(\ref{ebh}), and the generalization of the discussion in~\cite{Chakraborty:2023zdd} to $\epsilon_\pm\neq 0$,
imply that black holes in a negatively deformed theory have a maximal energy,
\be\label{ebhmax}
E_{BH}^{max}=-{p\over\Psi R_x}~.
\ee
\item
Similarly, eq.~(\ref{especb}) with the $a,b$ and $c$ in~(\ref{a}),(\ref{bb}) and~(\ref{cb}), respectively,
imply that long-strings excitations with winding $w_x$, and $n_x=q_{L,R}=0$, have a maximal (real) total energy,
\be\label{etmax}
E_{\rm total}^{max}=-{w_x\over\Psi R_x}~,
\ee
for the negatively deformed theories,~(\ref{psineg}).
This is precisely the $w_x/p$-fraction of the maximal energy,~(\ref{ebhmax}), of the black string.
Recalling that the (deformed) BTZ black hole is formed by $p$ $F1$s,
and since a winding $w_x$ long string amounts to $w_x$ out of these $p$ $F1$s,
this suggests that such long strings cannot be excited beyond~(\ref{etmax}), in the negatively deformed theories.
\item
Our results apply also to the spectrum of long-strings excitations in superstring theory on deformed global $AdS_3\times S^1$,
by inserting
\be\label{global}
{r_5 M_{BTZ}\over p}=-{k\over 2}~,
\ee
everywhere above (as e.g. in~\cite{Chakraborty:2023mzc,Chakraborty:2024ugc}).
\item
When $\lambda=\epsilon_-\epsilon_+=0$,
the asymptotic regime is not a flat spacetime (with a linear dilaton).
Nevertheless, the results for the spectrum above apply to these cases as well (e.g. via continuation in~$(\lambda,\epsilon_\pm)$).
\item
The manipulations above can be generalized straightforwardly to the superstring theory on $(\lambda,\epsilon_\pm)$-deformed
$AdS_3\times S^3$, and other $AdS_3\times G/H$ examples of the kind summarized e.g. in~\cite{Giveon:2003ku}, namely,
to a deformation that involves an internal affine $U(1)_L\times U(1)_R$ symmetry, regardless of other properties of internal space.
\item
In particular, it can also be generalized straightforwardly to cases in which the $J$ and $\bar J$ above
are associated with different Abelian isometries of the internal geometry, as e.g. in appendix B of~\cite{Chakraborty:2019mdf}.
\item
In the general $(\lambda,\epsilon_\pm)$-parameters space, e.g. when $\epsilon_\pm=0$ and $\lambda<0$,
the $p$ strings on $S^1_x$ that form the background should be regarded as
`negative strings,'~\cite{Chakraborty:2020swe,Giveon:2023rsk,Giveon:2024sgz};
the results above
(in the way presented in~(\ref{especb}) via the $a,b$ and $c$ in~(\ref{a}),(\ref{bb}) and~(\ref{cb}), respectively)
are blind to the signs of $(\lambda,\epsilon_\pm)$, and thus apply to the negative strings cases as well.
\end{enumerate}

\section{Summary}

A short summary of this note is:
\begin{itemize}
\item
We considered supersting theory on deformed BTZ $\times S^1\times{\cal N}$,
with the deformed BTZ $\times S^1$ being a black string (whose geometry is that of~\cite{Apolo:2021wcn}).~\footnote{In
the non-rotating, uncharged case.}
\item
We explained how to derive the spectrum of long-strings excitations in this string theory,
from the asymptotic value of the metric, $B$-field and linear-dilaton slope
(in sections 2 and 3).
\item
For a particular asymptotic value of the $B$-field, the long-strings spectrum
is consistent with the picture that a long-string excitation with winding $w$ in the undeformed BTZ worldsheet CFT,
namely, an excitation with $w_x=w$ in the asymptotically linear-dilaton deformed theory in sections 2 and 3,
has the following properties:
\begin{enumerate}
\item
A long-string excitation with $w=1$ deforms together with its contribution to the black-string
background as a block ${\cal M}$ of a symmetric product, ${\cal M}^p/S_p$,
where ${\cal M}$ is a $T\bar T+J\bar T+T\bar J$ deformed $CFT_2$, whose central charge is $c=6k$,
with $\sqrt{\alpha' k}$ being the curvature length of the undeformed BTZ geometry,
and $p$ is the number of $F1$s that form the (deformed) BTZ background.
\item
The energy of an emitted long string with $w>1$ deforms together with the background as a $Z_w$ twisted sector of the symmetric product.
\end{enumerate}
\item
This is in harmony with the single-trace $T\bar T+J\bar T+T\bar J$ holography conjecture.
\end{itemize}

\vspace{10mm}

\section*{Acknowledgments}
We thank L.~Apolo and S.~Chakraborty for discussions.
This work was supported in part by the ISF (grant number 256/22).

\vspace{10mm}

\section*{Appendix A: $\epsilon_-=\epsilon_+\equiv \epsilon/2$ for $M_{BTZ}=0$ in detail}

In this appendix, for illustration, we present details of napkin calculations,
leading to the results in superstring theory on a symmetric $(\lambda,\epsilon)$-deformed massless BTZ black hole times a circle,
\be\label{elere}
\epsilon\equiv\epsilon_-+\epsilon_+~,\qquad \epsilon_--\epsilon_+=0~.
\ee
The detailed steps of the algebra for general $\epsilon_\pm$ and $M_{BTZ}$,
leading to the bottom lines of this note, is similar to those below.

The various matrices and functions that show up along the way are
\be\label{gije}
g_{ij}=\pmatrix{
 -1&0&0\cr
 0&r_x^2&r_x^2r_y\epsilon\cr
 0&r_x^2r_y\epsilon&r_x^2r_y^2\lambda
},
\qquad
(g^{-1})^{ij}=\pmatrix{
 -1&0&0\cr
 0&\lambda &-\epsilon/r_y\cr
 0&-\epsilon/r_y&1/r_y^2
},
\ee
\be\label{bije}
b_{ij}=\pmatrix{
0 & -r_x & -r_xr_y\epsilon\cr
r_x& 0& 0\cr
r_xr_y\epsilon & 0 &0
},
\ee
\be\label{gmbgb}
(g-bg^{-1}b)_{ij}=\pmatrix{
0 & 0 & 0\cr
0 & 0& 0\cr
0 & 0 & r_y^2
},
\ee
\be\label{bge}
b_{ik}(g^{-1})^{kj}=\pmatrix{
0 & -1/r_x & 0\cr
-r_x& 0& 0\cr
-r_xr_y\epsilon & 0 &0
},
\ee
\be\label{niwie}
n_i=(e,n_x,n_y)~,\qquad m^i=(0,w_x,w_y)~,
\ee
\be\label{ngn}
n_i(g^{-1})^{ij}n_j+m^i(g-bg^{-1}b)_{ij}m^j=-e^2+\lambda n_x^2-{2\epsilon n_xn_y\over r_y}+{n_y^2\over r_y^2}+{w_y^2r_y^2}~,
\ee
\be\label{mbgn}
m^ib_{ik}(g^{-1})^{kj}n_j=-(w_x+r_yw_y\epsilon)r_x e~,
\ee
from which one finds that the $(\Delta,\bar\Delta)$ in~(\ref{dpd})--(\ref{dmd})
are identical to those in equation~(5.27)~of~\cite{Chakraborty:2019mdf}~(CGK),
with the map~\footnote{Which
follows from the factor $R_{CGK}$ difference in our canonical choice (w.l.g.) of the $\gamma_1$ periodicity
(versus the generic, redundant one in~\cite{Chakraborty:2019mdf}).}
\be\label{mapcgk}
(ER)_{CGK}=r_x e~,\qquad \hat{\lambda}_{CGK}=2\lambda~,\qquad\hat{\epsilon}_{CGK}=\sqrt{2}\epsilon~,
\ee
and~\footnote{Due to $\sqrt{2/\alpha'}$ and minus factors in~(\ref{ppqq}) versus $(q_L,q_R)_{CGK}$ (in~(5.23) of~\cite{Chakraborty:2019mdf});
when $\epsilon_-\neq\epsilon_+$, we should also interchange the left and right-handed ones,
$(\hat\epsilon_\pm)_{CGK}=\sqrt{2}\epsilon_\mp$, accordingly.}
\be\label{qqcgk}
(q_L,q_R)_{CGK}=\sqrt{\alpha'\over 2}(-q_R,q_L)~,
\ee
as well as
\be\label{wncgk}
(w,n)_{CGK}\equiv(w_x,n_x)~.
\ee

\section*{Appendix B: $B_{tx}$ and $B_{xy}$ in detail}

In this appendix, for illustration, we present details of calculations leading to the result for $B_{tx}$ and $B_{xy}$ in~(\ref{answer}).
For that purpose, it is sufficient to present the analysis for
\be\label{nnw}
n_x=n_y=w_y=0~,
\ee
which is what we do below.

Using~(\ref{dpd}) with the metric~(\ref{gij}), the $n_i$ and $m^i$ in~(\ref{niwi}) with~(\ref{nnw}),
and a {\it general $B$-field}, in the on-shell condition~(\ref{onshellb}),
and taking into account that the $b_{ij}$ in~(\ref{dpd}) are written w.r.t.
compact coordinates normalized s.t. they have $2\pi$ periodicity, as in eq.~(2.4.11) of~\cite{Giveon:1994fu} that we used,
while the $B_{ij}$ in~(\ref{answer}) are written w.r.t. to the $t,x,y$ coordinates, whose normalization is in~(\ref{dsto}),
one finds:
\beq\label{onshelld}
-{\Psi\over 2}\left(r_xe+{w_xr_x\over\Psi}\left[r_x B_{tx}\Psi+B_{xy}(\epsilon_--\epsilon_+)\right]\right)^2
&+&\nonumber\\
+{w_x^2r_x^2\over 2}\left(1+{B_{xy}^2\over\Psi r_x^2}\right)
-{2j(j+1)\over k}+N_L+N_R-1&=&0~,
\eeq
where $r_x$, $e$ and $\Psi$ are given in~(\ref{gammas}),~(\ref{eee}) and~(\ref{defpsi}), respectively.

Now, requiring that the undeformed BTZ limit, $\lambda,\epsilon_\pm\to 0$, gives rise to~(\ref{calet}) with~(\ref{lsbtzb}),
for harmony with~\cite{Martinec:2023plo},
and demanding compatibility
of the long-strings spectrum in the $(\lambda,\epsilon_\pm)$-deformed theories
with the single-trace $T\bar T+J\bar T+T\bar J$ holography conjecture, namely, imposing the relation
\be\label{aeebec}
a(r_x e_{\rm total})^2+b(r_x e_{\rm total})+c=0~,
\ee
where~\footnote{See~(\ref{etotal}) with~(\ref{ebhm}), and (\ref{ebh}), for the definitions of $e_{\rm total}$, and $e_{BH}$.}
\be\label{etd}
e_{\rm total}=e+w_x{e_{BH}\over p}~,
\ee
with
\be\label{abcd}
a=-{\Psi\over 2}~,\qquad b=-w_x~,
\ee
and
\be\label{cd}
c=-{2j(j+1)\over k}+N_L+N_R-1=w_x\left({\cal E}+w_x{r_5 M_{BTZ}\over p}\right)~,
\ee
gives the $B_{tx}$ and $B_{xy}$ in~(\ref{answer}).

Finally, repeating the algebra for generic momentum in the $x$ direction, $n_x$,
as well as generic Narain momenta in the $y$ direction, $(q_L,q_R)$,
and requiring harmony with single-trace $T\bar T+J\bar T+T\bar J$ holography,
as well as the appropriate BTZ limit,
determines also the value of $B_{ty}$ in~(\ref{answer}), and gives rise to the results in section 3.

\end{document}